\documentclass[prd,showpacs,preprintnumbers,amsmath,amssymb]{revtex4-1}

\usepackage{graphicx}
\usepackage{dcolumn}
\usepackage{bm}
\usepackage{hyperref}

\def\be{\begin{equation}}
\def\ee{\end{equation}}
\def\bea{\begin{eqnarray}}
\def\eea{\end{eqnarray}}
\def\ba{\begin{array}}
\def\ea{\end{array}}
\usepackage{color}

\begin{document}

\preprint{UTHET-11-0801}

\title{Hair on near-extremal Reissner-Nordstr\"om AdS black holes}

\author{James Alsup}
\email{jalsup@umflint.edu}
\affiliation{Computer Science, Engineering and Physics Department, The University of Michigan-Flint, Flint, MI 48502-1907, USA.
}%

\author{George Siopsis}%
 \email{siopsis@tennessee.edu}
\author{Jason Therrien}%
 \email{jtherrie@utk.edu}
\affiliation{%
Department of Physics and Astronomy,
The University of Tennessee,
Knoxville, TN 37996 - 1200, USA.
}%

\date{June 2012}

\begin{abstract}
We discuss hairy black hole solutions with scalar hair of scaling dimension $\Delta$ and (small) electromagnetic coupling $q^2$, near extremality. Using trial functions, we show that hair forms below a critical temperature $T_c$ in the region of parameter space $( \Delta ,q^2)$ above a critical line $q_c^2 (\Delta)$. For $\Delta > \Delta_0$, the critical coupling $q_c^2$ is determined by the AdS$_2$ geometry of the horizon. For $\Delta < \Delta_0$, $q_c^2$ is {\em below} the value suggested by the near horizon geometry at extremality. We provide an analytic estimate of $\Delta_0$ (numerically, $\Delta_0 \approx 0.64$). We also compute analytically the true critical line for the entire range of the scaling dimension. In particular for $q=0$, we obtain an instability down to the unitarity bound.
We perform explicit analytic calculations of $T_c$, the condensate and the conductivity.
We show that the energy gap in units of $T_c$ diverges as we approach the critical line ($T_c \to 0$).
\end{abstract}

\maketitle

\section{Introduction}
In recent years it has been shown that one can create a condensed matter/gauge theory duality using the AdS/CFT correspondence \cite{Maldacena:1997re}. This duality can give us insight in how a strongly coupled system behaves by reformulating the theory in terms of a weakly coupled theory of gravity. As of now the Hall effect \cite{Hartnoll:2007ai}, Nernst effect \cite{Hartnoll:2007ih}, and superfluidity/superconductivity \cite{Gubser:2008px,Hartnoll:2008vx} have been explored. Some reviews are \cite{Hartnoll:2009sz,Horowitz:2010gk,Herzog:2009xv}.

Most of the work has been done with superconductivity. Initially the probe limit was explored, in which the chemical potential $\mu\to 0$ whereas the charge of the scalar hair $q\to\infty$ so that the product $q\mu$ remains finite \cite{Hartnoll:2008vx}. This work was later extended to include the full backreaction on the geometry of the system \cite{Hartnoll:2008kx}.  Magnetic fields have also been added and found to exhibit the behavior of Type-II superconductors \cite{Albash:2008eh}. Other gravitational duals have been able to describe systems exhibiting striped superconductivity (see, e.g., \cite{Flauger:2010tv}). All dual systems show hallmarks of the strongly coupled, high $T_c$ superconductors. 

In this paper we explore the properties of the system at the opposite end of the probe limit at which the charge of the scalar hair $q$ is close to zero and the critical temperature $T_c$ is small (small $T_c/\mu$ limit) corresponding to a near extremal black hole.
In \cite{Hartnoll:2008kx}, Hartnoll, {\em et al.}, discovered that a phase transition is allowed even for $q=0$ (see also discussion in \cite{Denef:2009tp,Horowitz:2009ij}). It is not clear what the physical mechanism responsible for the condensation is near $q=0$, and whether it is related to the one in the probe limit ($q\to\infty$).
To address these questions, we explore the near extremal limit by performing analytic calculations of the properties of the system. There is a critical line in the two-dimensional parameter space $(\Delta , q^2)$
defined by \cite{Horowitz:2009ij}
\be\label{eqcr} q^2 = \frac{3+2\Delta (\Delta -3)}{4} ~,\ee
which is determined by the AdS$_2$ geometry of the horizon at extremality. Instability is expected above this line, with the critical temperature approaching zero as one approaches the critical line. By using trial functions, we show that indeed, there is an instability above the critical line \eqref{eqcr}, but the critical temperature vanishes on the line only for $\Delta \ge  \Delta_0$. We compute the scaling dimension $\Delta_0$ analytically (numerically, $\Delta_0 \approx 0.64$). For $\Delta < \Delta_0$, the critical temperature does not vanish on the line \eqref{eqcr} and there is a phase transition below this line. We find an analytic expression for the true critical line (where $T_c=0$) for $\Delta < \Delta_0$. Interestingly, for $q=0$, we obtain instability down to the minimum value $\Delta =0.5$ determined by unitarity (at $\Delta =0.5$, we obtain $T_c=0$).

We also calculate analytically the energy gap
\be\label{eqEg} E_g = \frac{\langle \mathcal{O}_\Delta \rangle^{1/\Delta}}{T_c} ~,\ee
which is found to diverge as we approach the critical line ($T_c\to 0$). A calculation of the conductivity (both analytic and numerical) shows that the profile of the conductivity exhibits a universal behavior near criticality.

In detail, in section \ref{sec2} we set up the system and fix the notation. In section \ref{sec3}, we determine the critical temperature analytically. In section \ref{sec4} we study the system below the critical temperature and calculate the energy gap both analytically and numerically. In section \ref{sec5} we calculate the conductivity analytically near criticality and compare with numerical results. Finally, in section \ref{sec6} we summarize our conclusions.

\section{Equations of motion}
\label{sec2}

We are interested in the dynamics of a scalar field of mass $m$ and electric charge $q$ coupled to a $U(1)$ vector potential in the dynamical backgound of a $3+1 -$~dimensional AdS black hole. The action is
\be\label{eqS1} S = \int d^4 x \sqrt{-g} \left[ \frac{R + 6/L^2}{16\pi G} - \frac{1}{4} F^{\mu\nu} F_{\mu\nu} - |D_\mu \phi |^2 - m^2 |\phi|^2 \right], \ee
where $D_\mu = \partial_\mu -iq A_\mu$.
For simplicity, we shall work with units in which $L=1$ and $16\pi G =1$.

It is convenient to write the complex scalar field in terms of real scalar fields $\Psi$ and $\theta$ as
\be \phi = \Psi\, e^{iq\theta}  . \ee
Then the action reads
\be\label{eqS3} S = \int d^4 x \sqrt{-g} \left[ R + 6 - \frac{1}{4} F^{\mu\nu} F_{\mu\nu} - \partial_\mu\Psi \partial^\mu\Psi - q^2 \Psi^2 (\partial_\mu\theta - A_\mu)^2 - m^2 \Psi^2 \right] . \ee
The phase $\theta$ plays the role of a St\"uckelberg field giving mass the the vector potential when the real scalar $\psi$ condenses. The local $U(1)$ transformation reads
\be A_\mu \to A_\mu + \partial_\mu\omega \ \ , \ \ \ \ \theta\to \theta + \omega . \ee
We can fix the gauge by setting $\theta = 0$.
The form of the action (\ref{eqS3}) is more general than (\ref{eqS1}), because in the former, we can continue $q^2$ to negative values. This leads to a negative squared mass for the vector potential which signals an instability in flat space. However, in AdS space negative squared masses are allowed as long as they are above the BF bound. We shall allow $q^2$ to take negative values and find that there is a lower bound 
for a phase transition to occur.

To solve the equations of motion, consider a black hole with planar horizon and metric
\be ds^2 = \frac{1}{z^2} \left[ - g(z) e^{-\chi (z)} dt^2 + 
d{\vec x}^2
+ \frac{dz^2}{g(z)} \right] .
\ee
We have chosen the coordinate $z$ so that the horizon is at $z=1$ ($g(1)=0$).
The system is kept under fixed charge density $\varrho$. We shall denote by $\rho$ the value of the charge density in units in which the radius of the horizon is 1.

All physical quantities are measured in units of the horizon radius, which hides an arbitrary scale in the system. This is remedied by reporting on scale-invariant (dimensionless) quantities only, as we shall be doing.

Assuming that the scalar field is of the form $\Psi (z)$ and the potential is an electrostatic scalar potential $A_0 = \Phi (z)$, the
equations of motion are \cite{Hartnoll:2008kx}
\bea\label{eq1}
\Psi'' + \left[ \frac{g'}{g} - \frac{\chi'}{2} - \frac{2}{z} \right] \Psi' + \left[ \frac{q^2\Phi^2 e^{\chi}}{g^2} - \frac{m^2}{z^2 g} \right] \Psi &=& 0 , \nonumber\\
\Phi'' + \frac{\chi'}{2} \Phi' - \frac{2q^2\Psi^2}{z^2 g} \Phi &=& 0,\nonumber\\
-\chi' + z{\Psi'}^2 + \frac{zq^2\Phi^2\Psi^2}{g^2} e^\chi &=& 0,\nonumber\\
\frac{g}{2} {\Psi'}^2 + \frac{z^2}{4} {\Phi'}^2 e^\chi - \frac{g'}{z} + \frac{3(g-1)}{z^2 } + \frac{m^2\Psi^2}{2z^2 } + \frac{q^2 \Psi^2 \Phi^2 e^\chi}{2g} &=& 0,
\eea
where prime denotes differentiation with respect to $z$
and to be solved in the interval $(0,1)$, where $z=1$ is the horizon and $z=0$ is the boundary.

Near the boundary ($z\to 0$), we have $g\to 1$, $\chi \to 0$ and so approximately
\be\label{eq3} \Psi \approx \Psi^{(\pm)} z^{\Delta_\pm} \ \ , \ \ \ \ \Phi \approx \mu - \rho z ,\ee
where
\be \Delta_\pm = \frac{3}{2} \pm \sqrt{\frac{9}{4} + m^2} ~.\ee
For $m^2 < - \frac{9}{4}$ (Breitenlohner-Freedman (BF) bound \cite{Breitenlohner:1982jf}), $\Delta_\pm$ have an imaginary part and the system is unstable.
For $-\frac{9}{4} \le m^2 < - \frac{5}{4}$, both modes are normalizable.
While a linear combination of these modes is allowed by the equations of motion, it turns out that any such combination is unstable \cite{Hertog:2004bb}.
However, if the horizon has negative curvature, such linear combinations lead to stable configurations in certain cases \cite{papa1}.
Thus, the system is labeled uniquely by the dimension $\Delta = \Delta_\pm$.
For $m^2 \ge - \frac{5}{4}$, only the mode of scaling dimension $\Delta_+$ is normalizable.
It follows that $\Delta = \Delta_- > \frac{1}{2}$ (unitarity bound).
 
Demanding at the horizon
\be\label{eq7} \Phi (1) = 0, \ee
$\mu$ is interpreted as the chemical potential of the dual theory on the boundary,
$\rho$ is the charge density on the boundary, and the leading coefficient in the expansion of the scalar yields vacuum expectation values of operators of dimension $\Delta_\pm$,
\be\label{eq4} \langle \mathcal{O}_{\Delta_\pm} \rangle = \sqrt 2 \Psi^{(\pm)} . \ee
The Hawking temperature is
\be\label{eqH}
\frac{T}{\sqrt{\varrho}} = - \frac{g'(1)}{4\pi\sqrt{-\Phi'(0)}} e^{-\chi (1)/2} \ . \ee
The equations of motion admit non-vanishing solutions for the scalar below a critical temperature $T_c$ where these operators condense.

Above the critical temperature ($T\ge T_c$), we have $\Psi =0$ and $\chi = 0$.
The equations of motion reduce to
\bea
\Phi'' &=& 0 ,\nonumber\\
\frac{z^2}{4} {\Phi'}^2 - \frac{g'}{z} + \frac{3(g-1)}{z^2 } &=& 0 .
\eea
which yield the Reissner-Nordstr\"om black hole
\be\label{eq8} g(z) = 1- \left( 1 + \frac{\rho^2}{4} \right) z^3 + \frac{\rho^2}{4} z^4 \ \ , \ \ \ \ \Phi (z) = \rho (1-z) , \ee
whose Hawking temperature is
\be\label{eq9} \frac{T}{\sqrt{\varrho}} = \frac{1}{4\pi\sqrt{\rho}} \left( 3- \frac{\rho^2}{4} \right) . \ee

\section{The critical temperature}
\label{sec3}

To find
the critical temperature $T_c$, we note that at $T=T_c$ the equation of motion for the scalar field $\Psi$ decouples from the rest of the equations of motion, because $\Psi$ can be treated as a perturbation near the phase transition, and consequently, $g$ and $\Phi$ are simply given by \eqref{eq8}. Therefore, to find $T_c$, we need to solve the equation of motion for $\Psi$, \be\label{eq14a} \Psi'' + \left[ \frac{g'}{g} - \frac{2}{z} \right] \Psi' + \left[ \frac{q^2\Phi^2 }{g^2} - \frac{m^2}{z^2 g} \right] \Psi = 0~, \ee
together with the boundary condition $\Psi \sim z^\Delta$ at the boundary ($z=0$)
in the Reissner-Nordstr\"om black hole background (\ref{eq8}).
 We will do so by working very close to the zero temperature limit of the black hole.  At extremality we have
\be \frac{\rho^2}{4} = 3 \ \ , \ \ \ \ T = 0 ~. \ee
Near the horizon ($z\to 1$), the scalar field behaves as
\be\label{eq17} \Psi \sim (1-z)^{\delta_\pm} \ \ , \ \ \ \ \delta_\pm = - \frac{1}{2} \pm \frac{i}{\sqrt{3}} \sqrt{q^2 - \frac{3+2\Delta(\Delta -3)}{4}} ~.\ee
This suggests that there is a potential instability above a critical value of the electromagnetic coupling constant, given by the critical line \eqref{eqcr} in the two-dimensional parameter space $(\Delta, q^2)$,
since both scaling dimensions $\delta_\pm$ are complex and therefore no solution is regular at the horizon. We shall show that this instability does indeed occur and calculate $T_c$ analytically.

This instability has a geometric origin.
Near the horizon the geometry is AdS$_2 \times \mathbb{R}^2$ with radius $L_2 = 1/\sqrt{6}$, and the scalar has an effective mass given by
\be m_{\mathrm{eff}}^2 = m^2 + g^{tt} q^2 \Phi^2 = m^2-2q^2. \ee
The system is unstable below the BF bound for AdS$_2$ space, $m_{\mathrm{eff}}^2 L_2^2 < - \frac{1}{4}$, which is in accord with \eqref{eqcr}.

One expects $T_c\to 0$ as we approach the critical line \eqref{eqcr} from above, whereas below the line no instability is expected. We shall show that this only partly true. For $\Delta \ge \Delta_0$, where $\Delta_0 \approx 0.64$ (we also compute $\Delta_0$ analytically), we obtain $T_c=0$ along the line \eqref{eqcr}. However, for $\Delta < \Delta_0$, $T_c\ne 0$ along this line and we obtain instability below this line down to a true critical line (on which $T_c=0$) that we compute analytically. Among other things, this implies that for $q=0$ we have instability down to the unitarity bound $\Delta = 0.5$.

\subsection{$m^2 =-2$}

First, we consider the special case $m^2 = -2$, in which $\Delta = 1,2$, and a solution can be built on hypergeometric functions. A similar approach for other values of $m^2$ yields Heun functions which are not as manageable. For the more general case, we shall rely on trial functions.

The critical temperature corresponds to a black hole near extremality, so we set
\be\label{eq14e} \frac{\rho^2}{4} = 3 - \epsilon  \ \ , \ \ \ \ T = T_0 = \frac{\epsilon}{4\pi} ~, \ee
where $\epsilon \lesssim 1$.
The critical temperature is given by
\be\label{eq14}
\frac{T_c}{\sqrt{\varrho}} = \frac{T_0}{\sqrt{\rho}} \approx \frac{\epsilon}{4\pi \, 12^{1/4}} ~.
\ee
To find the critical temperature (i.e., $\epsilon$), we need to solve the wave equation \eqref{eq14a} near extremality.
It is a second order equation for a real function, therefore the general solution is given in terms of two arbitrary {\em real} integration constants.
There are two boundary conditions to be satisfied, one at $z=0$, where $\Psi \sim z^\Delta$ ($\Delta =1,2$), and regularity at the horizon ($z=1$). These two fix one of the integration constants,
as well as the eigenvalue $\epsilon$. The second integration constant (an overall normalization) remains arbitrary, because the wave equation is linear.

To solve the wave equation, split the interval $[0,1]$ into two overlapping regions, one away from the horizon, defined by $1-z \gtrsim \epsilon$, and one near the horizon, defined by $1-z \lesssim \sqrt{\epsilon}$.
These two regions overlap for $\epsilon\lesssim 1-z \lesssim \sqrt{\epsilon}$, if $\epsilon\lesssim 1$.

Away from the horizon ($1-z\gtrsim \epsilon$), we may approximate the metric by the extremal metric.
Then we obtain the solution at the critical temperature away from the horizon,
\be\label{eqfar} \Psi = \Psi_{\mathrm{far}} = \mathcal{C} \frac{z}{z-z_0} \left( \frac{z-z_0^*}{z-z_0} \right)^{\frac{2\sqrt{2}-i}{2\sqrt{3}} q} \left( \frac{1-z}{z-z_0} \right)^{\delta_+} F \left( -\delta_- + 2\sqrt{\frac{2}{3}} q, -\delta_- - \frac{iq}{\sqrt{3}} ;
-2\delta_- ; 2z_0^2\frac{1-z}{z-z_0} \right) + \mathrm{c.c}~,\ee
where $z_0, z_0^*$ are the two complex roots of $g(z)$ with positive and negative imaginary part, respectively, and
$\delta_\pm$ are the scaling dimensions \eqref{eq17} with $m^2=-2$.
The multiplicative constant $\mathcal{C}$ is complex. We shall use the boundary conditions to determine its phase, whereas $|\mathcal{C}|$ will remain arbitrary.

Near the horizon ($1-z \lesssim \sqrt{\epsilon}$), we need to exercise care.
Let us perform the coordinate transformation
\be\label{eqzz} z = 1- \frac{\epsilon}{6} \zeta~, \ee
so that $\zeta \lesssim 1/\sqrt{\epsilon}$. Notice that $\zeta$ can be large, and the horizon is at $\zeta =0$.

Expanding around the horizon,
\be g(z) = -\frac{\epsilon}{6} \zeta g'(1) + \frac{1}{2} \left( -\frac{\epsilon}{6} \zeta \right)^2 g''(1) + \dots = \frac{\epsilon^2}{6} \zeta (1+\zeta) + \dots ~,\ee
where we used $g'(1)=-4\pi T$, $g''(1) = 6 \left( \frac{\rho^2}{4} -1 \right)$, and eq.\ \eqref{eq14e},
the metric becomes
\be ds^2 = \frac{1}{6} \left[ - \epsilon^2 \zeta (1+\zeta) dt^2 + \frac{d\zeta^2}{\zeta (1+\zeta)} \right] + {d\vec x}^2 + \dots ,\ee
where we omitted higher-order terms in $\epsilon$.
One can verify that the Hawking temperature is still given by (\ref{eq14}).
The electrostatic potential reads
\be \Phi = \frac{\rho\epsilon}{6} \zeta + \dots ~,\ee
and expanding in $\epsilon$, the wave equation for the scalar field at the critical temperature near the horizon becomes
\be\label{eq19} \zeta (1+\zeta) \Psi'' + \left( 2\zeta + 1 \right) \Psi' + \frac{1}{3} \left[ 1 + q^2 \frac{\zeta}{1+\zeta} \right] \Psi = 0~ ,\ee
where prime denotes differentiation with respect to $\zeta$.
The acceptable solution at the critical temperature near the horizon is
\be\label{eqnear} \Psi_{\mathrm{near}} (\zeta) = A\left( 1+\zeta \right)^{-iq/\sqrt{3}} F \left( -\delta_+ - \frac{iq}{\sqrt{3}} , -\delta_- - \frac{iq}{\sqrt{3}} ; 1; - \zeta \right), \ee
given in terms of a different hypergeometric function, where $A$ is a {\em real} arbitrary normalization constant. The other (complex) normalization constant $\mathcal{C}$ will be related to $A$  by matching \eqref{eqnear} with the solution \eqref{eqfar} far away from the horizon.
Notice that the other solution to \eqref{eq19} is discarded because it has a logarithmic singularity at the horizon.

It is easily deduced from the identity
\be F(\alpha,\beta;\gamma;x) = (1-x)^{\gamma-\alpha-\beta} F(\gamma-\alpha, \gamma-\beta; \gamma; x)  \ee
that $\Psi_{\mathrm{near}}$ is real,
\be \Psi_{\mathrm{near}} = A\left( 1+\zeta \right)^{iq/\sqrt{3}} F \left( -\delta_- + \frac{iq}{\sqrt{3}} , -\delta_+ + \frac{iq}{\sqrt{3}} ; 1; - \zeta \right)  = \Psi_{\mathrm{near}}^* ~. \ee
Next, we match the two solutions in the overlap region.

We approach this region from near the horizon for $\zeta\gg 1$.
The wave function becomes
\be\label{eq22} \Psi_{\mathrm{near}} \approx A\frac{\Gamma (-1-2\delta_-)}{\Gamma (-\delta_- - \frac{iq}{\sqrt{3}} )\Gamma (-\delta_- + \frac{iq}{\sqrt{3}} )} \zeta^{\delta_+} + \mathrm{c.c.},
\ee
In the far region, we need to take the limit $z\to 1$. The corresponding wave function (\ref{eqfar}) becomes
\be \Psi_{\mathrm{far}} \sim \mathcal{C} (1-z_0)^{\delta_-} \left( \frac{1-z_0^*}{1-z_0} \right)^{\frac{2\sqrt{2} -i}{2\sqrt{3}} q} (1-z)^{\delta_+} + \mathrm{c.c.}. \ee
Using (\ref{eqzz}), we deduce the relation
\be\label{eqCA} \mathcal{C} = A\frac{\Gamma (-1-2\delta_-)}{\Gamma (-\delta_- + \frac{iq}{\sqrt{3}} ) \Gamma (-\delta_- - \frac{iq}{\sqrt{3}} )}
\left( \frac{\epsilon}{6} \right)^{-\delta_+} (1-z_0)^{-\delta_-} \left( \frac{1-z_0}{1-z_0^*} \right)^{\frac{2\sqrt{2} -i}{2\sqrt{3}} q} ~, \ee
between the constants $A$, $\mathcal{C}$, and $\epsilon$.
Thus, the phase of $\mathcal{C}$ has been fixed by imposing one of the boundary conditions (regularity at the horizon). An overall normalization constant $A$ remains arbitrary, as expected, since the wave equation is linear.

\begin{figure}[t]
\includegraphics[width=.49\textwidth]{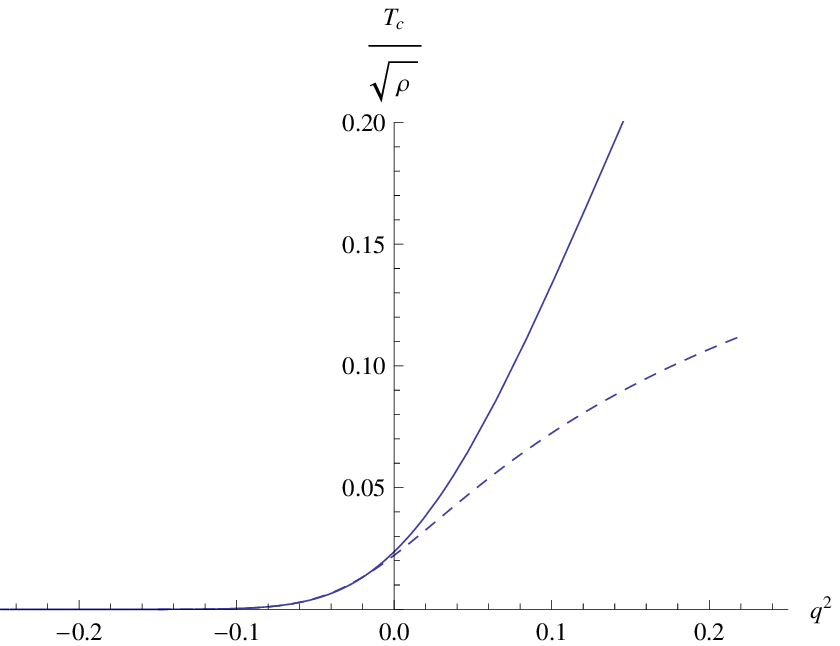}
\includegraphics[width=.49\textwidth]{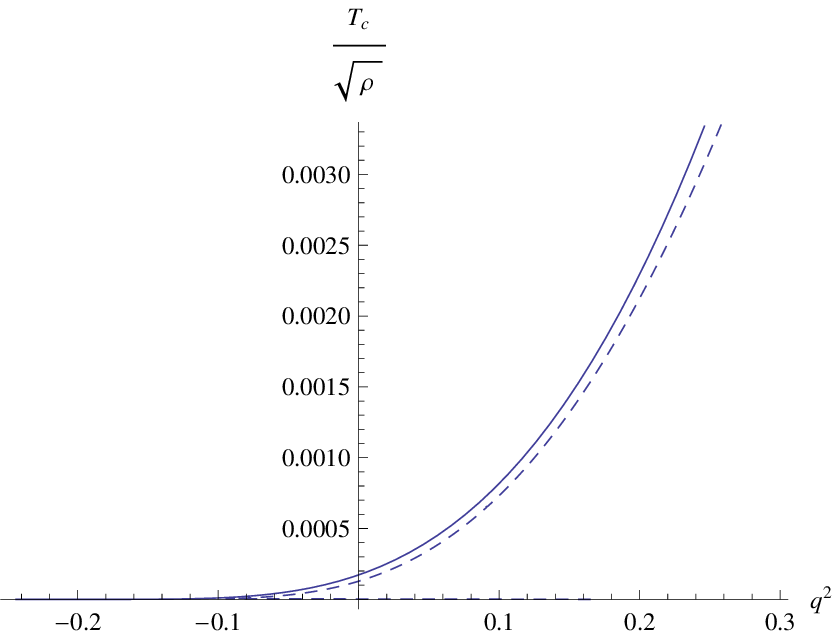}
\caption{The critical temperature $T_c$ as a function of $q^2$ found by a numerical solution (dashed) and an analytic one based on hypergeometric functions (solid) for $\Delta=1$ (left panel) and $\Delta=2$ (right panel).  $T_c \to 0 $ as $q^2 \to - \frac{1}{4}$.}
\label{fig1}
\end{figure}

The eigenvalue $\epsilon$ will be fixed by imposing the remaining boundary condition at $z=0$.
Focusing on $\Delta = \Delta_+=2$, we demand the asymptotics $\Psi \sim z^2$  near the boundary, giving
\be\label{eqfar1} \mathcal{C} (-z_0)^{\delta_-} \left( \frac{z_0^*}{z_0} \right)^{\frac{2\sqrt{2} -i}{2\sqrt{3}} q} F \left( -\delta_- + 2\sqrt{\frac{2}{3}} q , -\delta_- - \frac{iq}{\sqrt{3}} ;
- 2\delta_- ; -2z_0 \right)
+ \mathrm{c.c.}
= 0 ~.
\ee
By solving this constraint with $\mathcal{C}$ given in \eqref{eqCA} (notice that the arbitrary {\em real} normalization constant $A$ drops out), we obtain $\epsilon$, and therefore the critical temperature
\be\label{eqTrho} \frac{T_c}{\sqrt{\varrho}} = \frac{\epsilon}{4\pi 12^{1/4}} ~, \ee
as a function of the charge $q$.  The solution may be written in terms of the hypergeometric functions in \eqref{eqfar1}, but we will not present it here.  At the minimum value $q^2 = q_c^2 = - \frac{1}{4}$, found from the critical line \eqref{eqcr} with $m^2=-2$, we obtain $T_c =0$.
For $q^2 > q_c^2$, $T_c$ is an increasing function of $q^2$.  This behavior is shown in figure \ref{fig1}.

For $q^2=0$, we deduce from (\ref{eqfar1}),
\be\label{eqed2} \epsilon \approx 0.004 ~, \ee
and the critical temperature is
\be \frac{T_c}{\sqrt{\varrho}} \approx 1.7\times 10^{-4} ~. \ee
For the other boundary condition, $\Delta =1$, we have $\Psi \sim z + \mathcal{O} (z^3)$ as $z \to 0$. The calculation may be done in a similar manner, giving the value at $q^2=0$,
\be\label{eqed1}
\epsilon \approx .553 ~,
\ee
and the critical temperature
\be
\frac{T_c}{\sqrt{\varrho}} \approx .024~.
\ee
The critical temperature as a function of $q^2$ is shown in comparison with numerical results \cite{Hartnoll:2008kx} in figure \ref{fig1}.
It should be pointed out that our analytic results are valid in the small $q^2$ regime where $\epsilon$ is small, since we are performing a small $\epsilon$ expansion.
As $q^2$ increases, so does $\epsilon$, and so corrections from higher perturbative orders become important. Also, comparing different scaling dimensions, the corrections for $\Delta =1$ (eq.\ \eqref{eqed1}) are larger, because $\epsilon$ is not as small as for $\Delta =2$ (eq.\ \eqref{eqed2}). They become increasingly accurate as we approach the critical point $q_c^2 = - \frac{1}{4}$. In this limit, the numerical analysis becomes unstable.

\subsection{Variational method}

Next, we solve the wave equation and determine the critical temperature for arbitrary scaling dimension $\Delta$ using a variational method. Our results extend earlier work by Hartnoll, {\em et al.}\  \cite{Hartnoll:2008kx} who used a variational method to show that there was a phase transition in the case $m^2=-2$, $q^2=0$. Their trial functions worked well for $\Delta =\Delta_-$, however they were unable to find a trial function which would show instability in the case $\Delta = \Delta_+$ where the critical temperature is lower.
We find trial functions for all scaling dimensions $\Delta = \Delta_\pm$ and coupling constants $q^2$ that show that we have instability down to a critical line $q^2 = q_c^2 (\Delta)$ that we compute analytically.

Let
\be \Psi = \frac{\langle\mathcal{O}_\Delta\rangle}{\sqrt{2}} z^\Delta F(z) ~. \ee
The equation of motion for the scalar field \eqref{eq14a} in the background \eqref{eq8} reads
\be\label{eq40} -F''-\left[ \frac{g'}{g}+ \frac{2(\Delta -1)}{z} \right] F' + \frac{\Delta}{z^2} \left[ (\Delta -3) \left( \frac{1}{g} -1 \right) - z \frac{g'}{g} \right] F = q^2\rho^2 \frac{(1-z)^2}{g^2} F ~. \ee
For a given $\rho$, $q^2$ is an eigenvalue determining $T_c$.
It extremizes the expression
\be\label{eq41} q^2 = \frac{\mathcal{N}}{\mathcal{D}} ~, \ee
where
\be\label{eq41a} \mathcal{N} = \int_0^1 dz\ z^{2\Delta -2} [ g {F'}^2 + \Delta z (\Delta + \Delta \frac{\rho^2}{4} (1-z) - \frac{\rho^2}{4} z ) F^2 ] \ , \ \
\mathcal{D} = \rho^2\int_0^1 dz\ z^{2\Delta -2} \frac{(1-z)^2}{g} F^2 ~. \ee
Of the (generally infinite number) of extrema (eigenvalues), we need to select the minimum.
To this end, consider the one-parameter family of trial functions
\be\label{eq42} F_{\alpha} (z) = 1 - \frac{\alpha z^2}{\sqrt{z_--z}} ~, \ee
where $z_-$ is the inner horizon ($z_->z_+ =1$ and $g(z_-)=0$).

The critical point $q_c^2$ can be found by substituting the trial function \eqref{eq42} in \eqref{eq41a}, and taking the extremal limit $\rho^2 \to 12$. We obtain
\bea \mathcal{N} &=& -\frac{3-2\Delta(3-\Delta)}{2}\alpha^2 \ln (z_--1)+ n_0(\alpha) + \mathcal{O} (\sqrt{z_--1}) ~, \nonumber\\
\mathcal{D} &=& -2\alpha^2 \ln (z_--1)+ d_0(\alpha) + \mathcal{O} (\sqrt{z_--1}) ~.
\eea
The functions $n_0(\alpha)$ and $d_0(\alpha)$ can be found explicitly, but will not be needed for our purposes.

In the extremal limit $z_- \to 1$, we obtain from eq.\ \eqref{eq41},
\be\label{eq41b} q^2 = \frac{3 - 2\Delta (3-\Delta)}{4} + \mathcal{O} \left( \frac{1}{-\ln (z_--1)} \right) ~, \ee 
and therefore $q_c^2$, which is given by \eqref{eq41b} in the limit $z_-\to 1$, is in agreement with expectations from geometrical considerations at the horizon (eq.\ \eqref{eqcr}).

\begin{figure}[ht!]
\includegraphics[width=.8\textwidth]{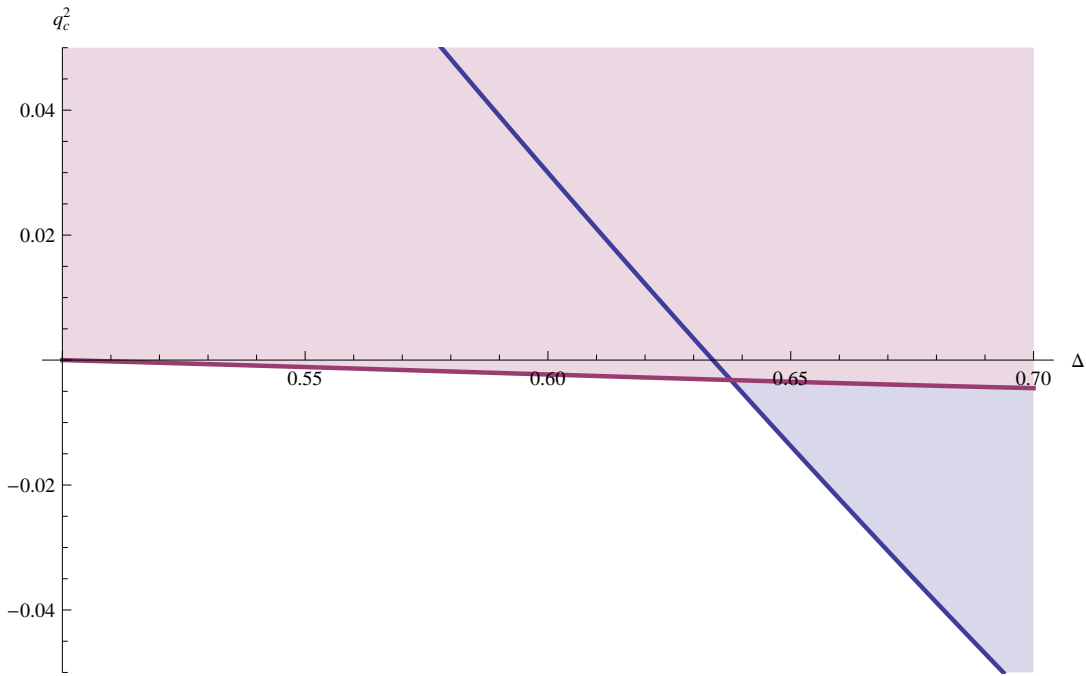}
\caption{The critical coupling constant $q_c^2$ {\em vs.}\ $\Delta$ (boundary of the shaded region). The curve close to the horizontal axis represents the result \eqref{eq41c} of our variational method, whereas the steeper curve is the result \eqref{eqcr} one obtains from geometrical considerations.}
\label{figqcr}
\end{figure}

\begin{figure}[b!]
\includegraphics[width=.8\textwidth]{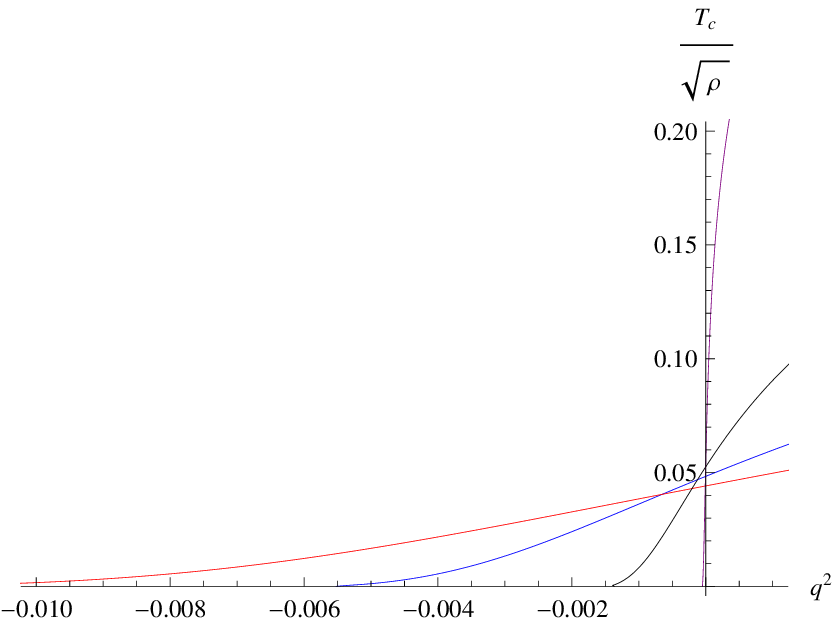}
\caption{The critical temperature {\em vs.}\ the coupling constant $q^2$ for $\Delta = 0.634, 0.578, 0.525, 0.501$ (curves become steeper as $\Delta$ decreases).}
\label{figq2}
\end{figure}

The above conclusion is valid as long as $\alpha\ne 0$. In general, setting $\alpha =0$, one obtains a value of the ratio \eqref{eq41} which is higher than the minimum, validating the above conclusion. However, for sufficiently small $\Delta$, the value of the ratio \eqref{eq41} at $\alpha=0$ is lower than the value \eqref{eq41b}. In this case, the minimum is attained at $\alpha=0$. We deduce the critical point (in the limit $z_-\to 1$),
\be\label{eq41c} q_c^2 = \frac{n_0(0)}{d_0(0)} = \frac{\Delta (1-\Delta )(2\Delta -1)}{4(2\Delta +1)}\, \frac{\Im z_0}{\Im \frac{1}{z_0} F( 2\Delta -1, 1; 2\Delta ; \frac{1}{z_0}) } ~. \ee
The two possible critical coupling constants $q_c^2$ are plotted in figure \ref{figqcr} as functions of the scaling dimension $\Delta$. The two critical lines meet at $\Delta_0$ which solves
\be \frac{3 - 2\Delta_0 (3-\Delta_0)}{4} = \frac{\Delta_0 (1-\Delta_0 )(2\Delta_0 -1)}{4(2\Delta_0 +1)}\, \frac{\Im z_0}{\Im \frac{1}{z_0} F( 2\Delta_0 -1, 1; 2\Delta_0 ; \frac{1}{z_0}) } ~. \ee
Numerically,
\be \Delta_0 \approx 0.64 ~. \ee
For $\Delta > \Delta_0$, the critical point is given by \eqref{eqcr}, whereas for $\Delta < \Delta_0$, it is given by \eqref{eq41c}. We have instability in the shaded region of figure \ref{figqcr} (above the minimum of the two critical lines for each $\Delta$). Notice that even in the range $\Delta < \Delta_0$, we have $q_c^2 < 0$, with $q_c^2=0$ at the unitarity bound $\Delta =0.5$. Therefore Reissner-Nordstr\"om black holes are unstable against neutral hair down to the unitarity bound, contrary to what one would expect by geometrical considerations at the horizon. This result is confirmed by a numerical calculation of the critical temperature as a function of $q^2$ for various values of $\Delta$ (figure \ref{figq2}).

The above results show conclusively that we have instability for all $q^2 > q_c^2$.
Indeed, for a given $q^2 > q_c^2$, define the action
\be S = \mathcal{N} - q^2 \mathcal{D} ~, \ee  
with $\mathcal{N}$ and $\mathcal{D}$ given in \eqref{eq41a}. The action $S$ is extremized ($\delta S = 0$) by the solutions of the wave equation \eqref{eq40}.
It becomes negative for trial functions corresponding to eigenvalues (coupling constants) $\tilde q^2 < q^2$, since
\be S = ( \tilde q^2 - q^2 ) \mathcal{D} < 0~. \ee
This shows that we have instability above the critical point ($q^2 > q_c^2$). By the same argument, below the critical point ($q^2 < q_c^2$), we have $S >0$, since $\tilde q^2 > q_c^2 > q^2$ for all trial functions, therefore no instability against scalar hair.

\begin{figure}
\includegraphics[width=.49\textwidth]{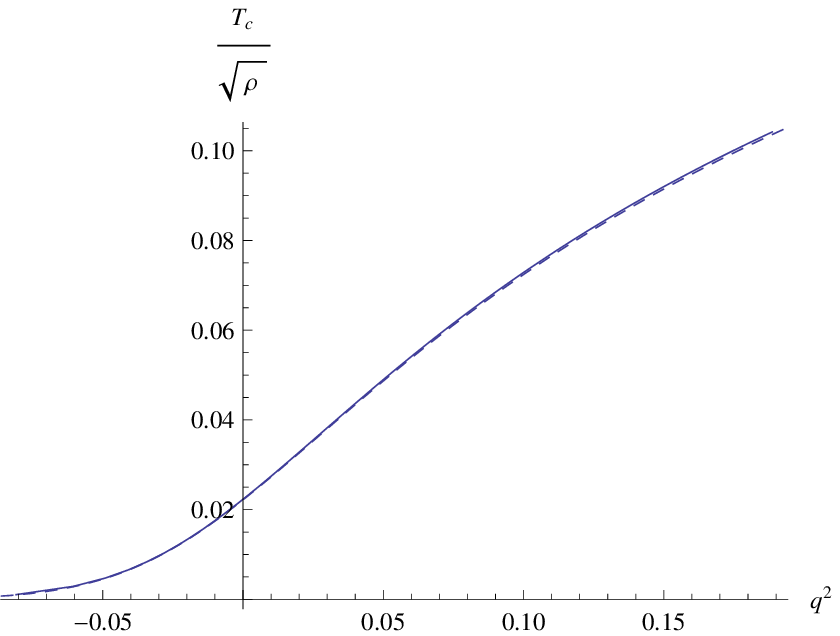}
\includegraphics[width=.49\textwidth]{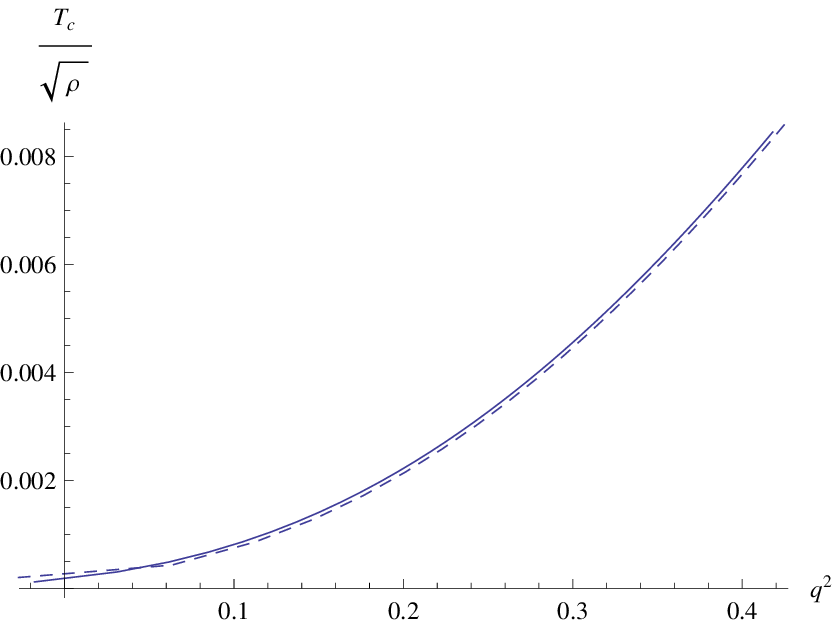}
\caption{The critical temperature $T_c$ found by a variational method (dashed) compared to numerical results (solid), for $\Delta=1$ (left panel) and $\Delta=2$ (right panel).}
\label{figvn}
\end{figure}

We obtain analytic expressions for the critical temperature as a function of $q^2$ for given $\Delta$ by using \eqref{eq9}. These expressions are compared with numerical data in figure \ref{figvn}. We obtain excellent agreement (the curves are almost indistinguishable) both near and  away from the critical point, as shown in figure \ref{figvn}.
The variational method is superior to the numerical solution as we approach the critical point ($T_c\to 0$), because the latter becomes unstable there.

\section{Below the critical temperature}
\label{sec4}

Below the critical temperature (\ref{eq14}), $\Psi$ does not vanish. We shall use perturbation theory to obtain solutions to the equations of motion. We expect the perturbative expansion to converge rapidly because the critical temperature is very small near extremality. We therefore expand
\bea\label{eqli}
\Psi &=& \langle\mathcal{O}_\Delta \rangle \Psi_0 + \langle\mathcal{O}_\Delta \rangle^3 \Psi_1 + \dots \nonumber\\
\Phi &=& \Phi_0 + \langle\mathcal{O}_\Delta \rangle^2 \Phi_1 + \dots\nonumber\\
\chi &=& \chi_0 + \langle\mathcal{O}_\Delta \rangle^2 \chi_1 + \dots\nonumber\\
g &=& g_0 + \langle\mathcal{O}_\Delta \rangle^2 g_1 + \dots
\eea
The charge density (in units in which the radius of the horizon is $z_+=1$) is similarly expanded,
\be \rho = -\Phi'(0) = \rho_0 + \langle\mathcal{O}_\Delta \rangle^2 \rho_1 + \dots \ee
where $\rho_0$ is the value of the charge density $\varrho$ in units of the radius of the horizon at the critical temperature.

The temperature of the hairy black hole \eqref{eqH}
is expanded as
\be\label{eqT2} \frac{T}{\sqrt{\varrho}} = \frac{T_0}{\sqrt{\rho_0}} \left[ 1 - \langle\mathcal{O}_\Delta \rangle^2 \mathcal{T}_1 + \dots \right] ~, \ee
around the critical temperature \eqref{eq14}, where
\be \mathcal{T}_1 = -\frac{g_1'(1)}{g_0'(1)} + \frac{1}{2} \chi_1 (1) +\frac{\rho_1}{2\rho_0} ~. \ee
At zeroth order, we have
\be\label{eq0} \Phi_0 = \rho_0 (1-z) \ , \ \ \chi_0 = 0 \ , \ \ g_0 = 1-\left( 1 + \frac{\rho_0^2}{4} \right) z^3+ \frac{\rho_0^2}{4} z^4~, \ee
and $\Psi_0$ obeys the wave equation in the background \eqref{eq0} and has already been calculated,
\be \Psi_0 = \frac{1}{\sqrt{2}} z^\Delta F(z) ~, \ee
where $F$ is normalized by $F(0)=1$.

At first order, we obtain
\be \chi_1 (z) = \int_0^z dz' \left[ z'(\Psi_0'(z'))^2 + q^2 \rho_0^2 \frac{z' (1-z')^2}{g_0^2(z')} (\Psi_0(z'))^2 \right] ~. \ee
We deduce the correction to the electric potential
\be \Phi_1 (z) = -\int_z^1 dz' \Phi_1'(z') \ , \ \ \Phi_1' (z) = -\rho_1 +\rho_0 \varphi(z) ~. \ee
The parameter $\rho_1$ is an integration constant to be determined,
and
\be \varphi (z) = \int_z^1 dz' \left[ \frac{z'}{2} (\Psi_0'(z'))^2
+ q^2 \left( 2\frac{1-z'}{{z'}^2 g_0(z')} + \rho_0^2 \frac{z'(1-z')^2}{2 g_0^2(z')} \right) (\Psi_0(z'))^2 \right] ~.
\ee
We also obtain
\be g_1(z) = z^3 \left[ -\frac{\rho_0\rho_1}{2} (1-z) +\mathcal{G} (z) \right] ~, \ee
where $\mathcal{G} (1) = 0$ and
\be \mathcal{G}'(z) = \left( \frac{m^2}{2{z}^4} + \frac{q^2 \Phi_0^2(z)}{2{z}^2 g_0(z)} \right) \Psi_0^2 (z) + \frac{g_0(z)}{2{z}^2} (\Psi_0'(z))^2 + \frac{\rho_0^2}{4} (\chi_1(z) -2\varphi (z)) ~. \ee
Therefore at the horizon,
\be g_1'(1) = \frac{\rho_0^2}{4} \chi_1(1) + \frac{m^2}{2} \Psi_0^2(1) + \frac{\rho_0\rho_1}{2} ~. \ee
The remaining unknown parameter $\rho_1$ is found from the first-order correction to the wave equation,
\be \Psi_1'' + \left[ \frac{g_0'}{g_0} - \frac{2}{z} \right] \Psi_1'+ \left[ \frac{q^2\Phi_0^2}{g_0^2} - \frac{m^2}{z^2g_0} \right] \Psi_1 = -\mathcal{H}_1 \Psi_0 ~, \ee
where
\be \mathcal{H}_1 \Psi_0 = \left[ \frac{g_1'}{g_0} - \frac{g_1g_0'}{g_0^2} - \frac{\chi_1'}{2} \right] \Psi_0' + \left[ \frac{m^2 g_1}{z^2 g_0^2} - \frac{2q^2\Phi_0 ( g_0\Phi_1-g_1\Phi_0)}{g_0^3} \right] \Psi_0 ~. \ee
After taking the inner product with $\Psi_0$ and using the zeroth-order wave equation, we arrive at the condition
\be \int_0^1 \frac{dz}{z^2} g_0(z) \Psi_0(z) \mathcal{H}_1 \Psi_0(z) = 0 ~, \ee
determining $\rho_1$. This is linear in $\rho_1$ and easily solved,
\be \rho_1 = \frac{\int_0^1 \frac{dz}{z^2} \left(\left[ (z^3\mathcal{G})' - \frac{z^3\mathcal{G}g_0'}{g_0} - g_0\frac{\chi_1'}{2} \right] \Psi_0\Psi_0' + \left[ \frac{m^2 z\mathcal{G}}{ g_0} + \frac{2q^2\rho_0^2 (1-z) ( g_0\varphi-z^3(1-z)\mathcal{G})}{g_0^2} \right] \Psi_0^2 \right)}{\rho_0\int_0^1 \frac{dz}{z^2} \left( \left[ \frac{z^2(3-4z)}{2} - \frac{g_0'z^3(1-z)}{2g_0} \right] \Psi_0\Psi_0' + \left[ \frac{m^2 z(1-z)}{2 g_0} - \frac{q^2 (1-z)^2( 2g_0+\rho_0^2z^3(1-z))}{g_0^2} \right] \Psi_0^2 \right)} ~. \ee
From the second-order expression \eqref{eqT2} for the temperature, we deduce the leading behavior of the condensate below the critical temperature,
\be \langle \mathcal{O}_\Delta \rangle = \frac{1}{\sqrt{\mathcal{T}_1}} \sqrt{1-\frac{T}{T_c} }~. \ee
By setting $T=0$, we deduce the first-order approximation to the energy gap,
\be \frac{\langle\mathcal{O}_\Delta \rangle^{1/\Delta}}{T_c} =
\frac{1}{T_0 \mathcal{T}_1^{1/(2\Delta)}} ~.\ee
\begin{figure}
\includegraphics[width=.31\textwidth]{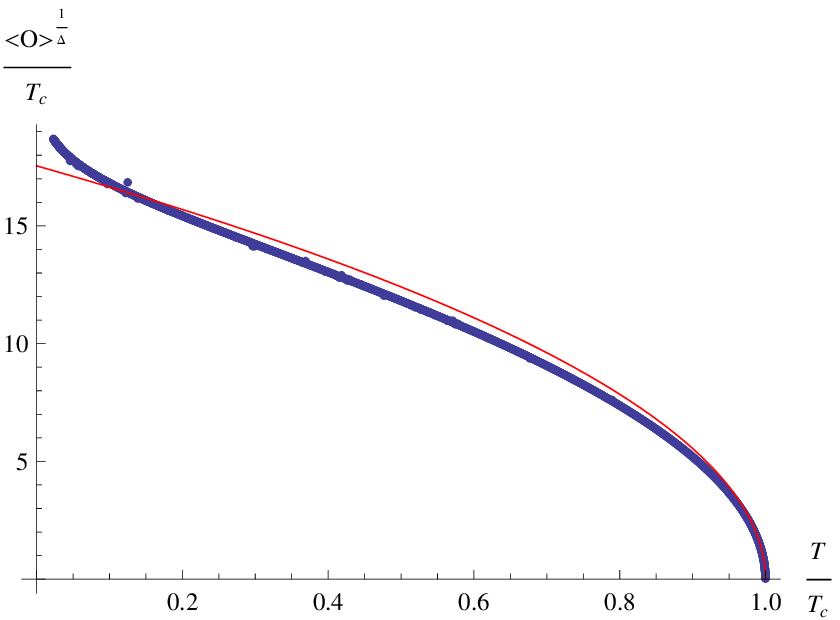}
\includegraphics[width=.31\textwidth]{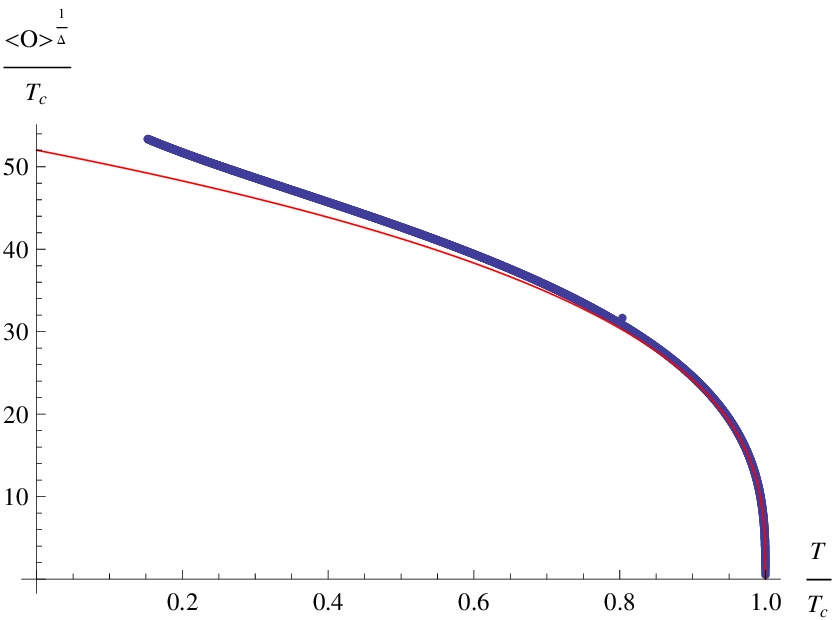}
\includegraphics[width=.31\textwidth]{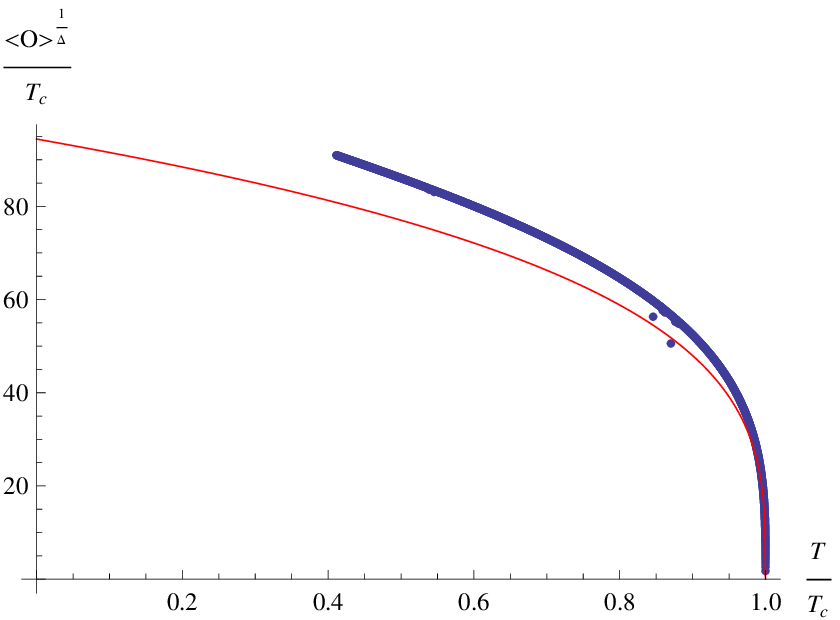}
\caption{The energy gap for $q^2=0$, and $\Delta=1,1.5,1.7$ (left to right). The thin line is our first-order analytic result, the thick one is from numerics.}
\label{fig4a}
\end{figure}

%


\begin{figure}[t]
\includegraphics[width=.49\textwidth]{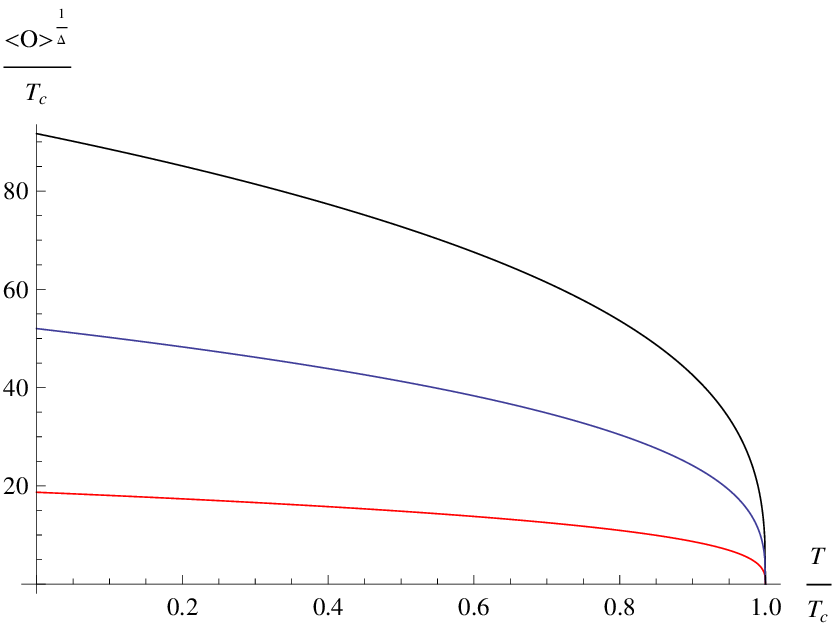}
\includegraphics[width=.49\textwidth]{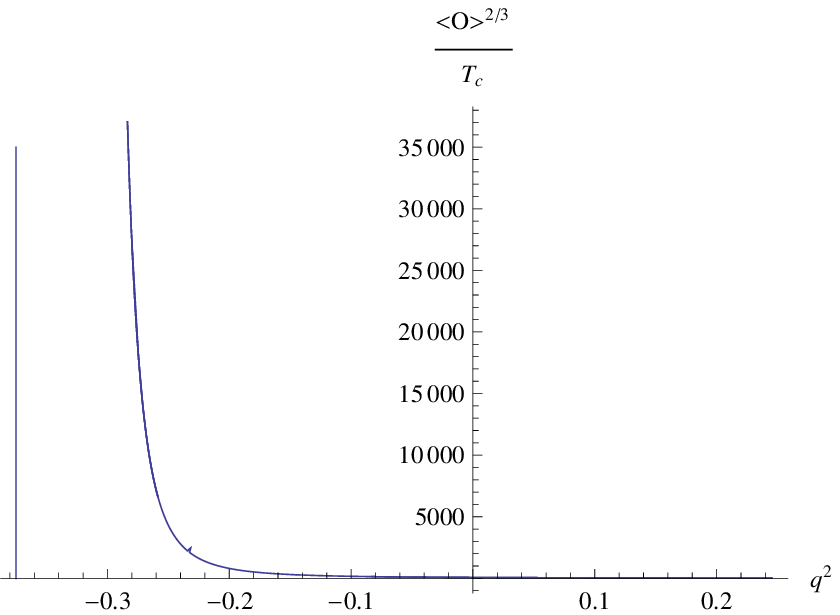}
\caption{Left panel: energy gap {\em vs} temperature for $\Delta=1.5$ $q^2=0.38$ (lower curve), $q^2=0$ (middle curve), and $q^2=-.125$ (upper curve).  Right panel: energy gap {\em vs} $q^2$ at temperature $T=.95 T_c$ (asymptote at $q^2 = -0.375$). } 
\label{fig4}
\end{figure}


In figure \ref{fig4a} we compare our first-order analytic results for the gap at $q^2 =0$ to numerical results obtained by solving the equations of motion numerically. The agreement improves as one decreases $q^2$, because the critical temperature $T_c$ decreases. In figure \ref{fig4}, we plot the gap as a function of $q^2$ and show that it diverges as $q^2\to q_c^2$ (i.e., $T_c\to 0$).


\section{Conductivity}
\label{sec5}
Next, we calculate the conductivity of the system.  To this end, we introduce perturbations for the electromagnetic gauge field while respecting translational invariance via
\be
A_x\sim g_{tx}\sim e^{-i\omega t}~.
\ee
The equations of motion are \cite{Hartnoll:2008kx}
\bea
\left(A_x'g e^{\frac{-\chi}{2}}\right)'+ \left( \frac{\omega^2 e^{\frac{\chi}{2}}}{g} - 2q^2\frac{\Psi^2}{z^2} e^{\frac{-\chi}{2}} \right) A_x &=& -\Phi'e^{\frac{\chi}{2}}\left(z^2g_{tx}'+2zg_{tx}\right) ~, \nonumber\\
2g_{tx}+z(\Phi' A_x+g_{tx}') &=& 0~,
\eea
which yield the equation for the electromagnetic perturbation (in Schr\"odinger form)
\be\label{eqwc} - \frac{d^2 A_x}{dz_*^2} + V A_x = \omega^2 A_x ~,\ee
where $z_*$ is a ``tortoise'' coordinate defined by
\be z_* = \int_0^z \frac{dz'}{g(z')} e^{\chi(z') /2} ~,\ee
with $z_* = 0$ at $z=0$,
and the potential is given by
\be\label{eqexpot} V = g  \left( 2q^2\frac{\Psi^2}{z^2} e^{-\chi} + z^2 {\Phi'}^2 \right) ~. \ee
The wave equation \eqref{eqwc} ought to be solved subject to ingoing boundary condition, $A_x\sim e^{i\omega z_*}$, at the horizon.  At the boundary, expanding the gauge field in a series,
\be
A_x\approx A_x^{(0)}+z A_x^{(1)}+\dots~,
\ee
we read off the conductivity,
\be\label{eq72}
\sigma=\frac{A_x^{(1)}}{i\omega A_x^{(0)}}~.
\ee
To visualize this, it is convenient to extend the domain to the entire real axis with $V=0$ for $z_*<0$.
Consider a wave of ``energy'' $\omega^2$ incident from the left, $A_x = e^{i\omega z_*}$ for $z_*<0$. At $z_*=0$, it gets partly reflected and partly transmitted, so that for $z_*<0$, $A_x = e^{i\omega z_*} + \mathcal{R} e^{-i\omega z_*}$, whereas for $z_*>0$, we have a transmitted wave which is purely ingoing at the horizon. The reflection coefficient is then related to the conductivity \eqref{eq72} by \cite{Horowitz:2009ij}
\be \sigma = \frac{1-\mathcal{R}}{1+\mathcal{R}} \ee
To find $\mathcal{R}$, write the wave equation in integral form as
\be A_x(z_*) = e^{i\omega z_*} - \int_0^\infty dz_*' G(z_*,z_*') V(z_*') A_x(z_*') ~,\ee
where the Green function satisfies
\be -\partial_{z_*}^2 G(z_*,z_*') - \omega^2 G(z_*,z_*') = \delta (z_*-z_*') ~.\ee
Explicitly,
\be G(z_*,z_*') = \frac{i}{2\omega} \left[ \theta (z_*'-z_*) e^{i\omega (z_*'-z_*)} + \theta(z_* -z_*') e^{-i\omega (z_*'-z_*)} \right] ~.\ee
The reflection coefficient is then
\be \mathcal{R} = - \int_0^\infty dz_* G(0,z_*) V(z_*) A_x(z_*) ~,\ee
and can be calculated perturbatively,
\be\label{eq78} \mathcal{R} = - \int_0^\infty dz_* G(0,z_*) V(z_*) e^{i\omega z_*} + \int_0^\infty dz_* G(0,z_*) V(z_*) \int_0^\infty dz_*' G(z_*,z_*')V(z_*')e^{i\omega z_*'} + \dots~. \ee
\begin{figure}[t]
\includegraphics[width=.49\textwidth]{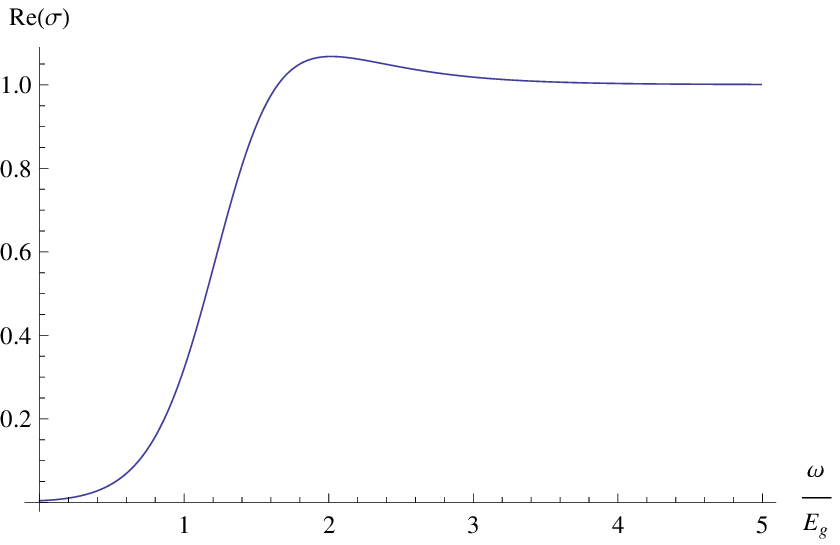}
\includegraphics[width=.49\textwidth]{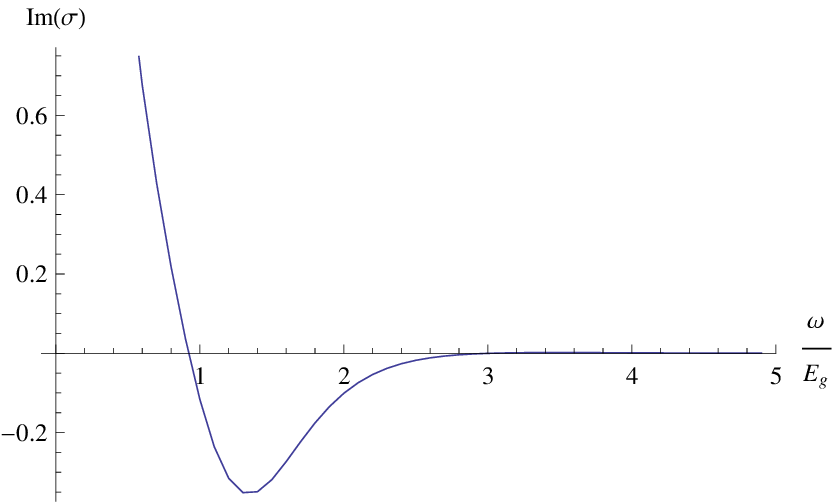}
\caption{Real and imaginary parts of the conductivity. } 
\label{figq0n}
\end{figure}
The numerical results are shown in figure \ref{figq0n}.
The analytic results at second order are compared with the numerical solution in figure \ref{figconda}. As expected, we have agreement at high $\omega$. As we go to higher perturbative orders, the agreement extends to a wider range of frequencies.
\begin{figure}[t]
\includegraphics[width=.49\textwidth]{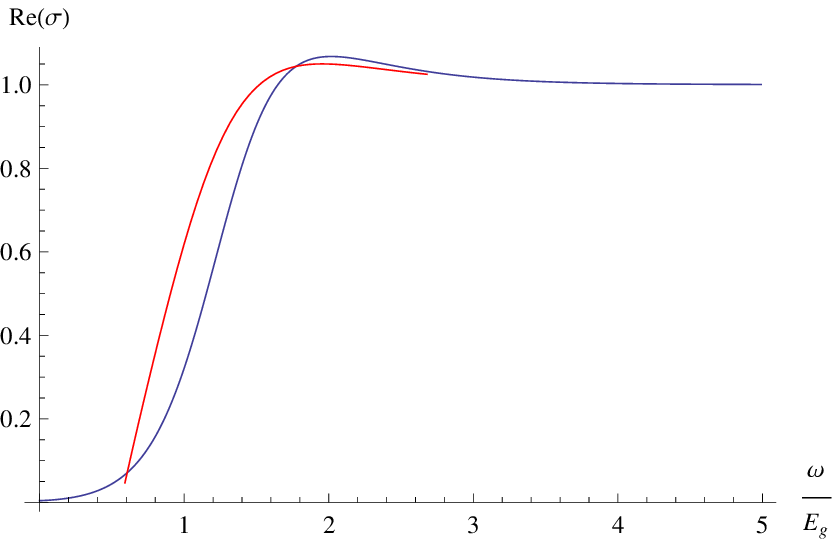}
\includegraphics[width=.49\textwidth]{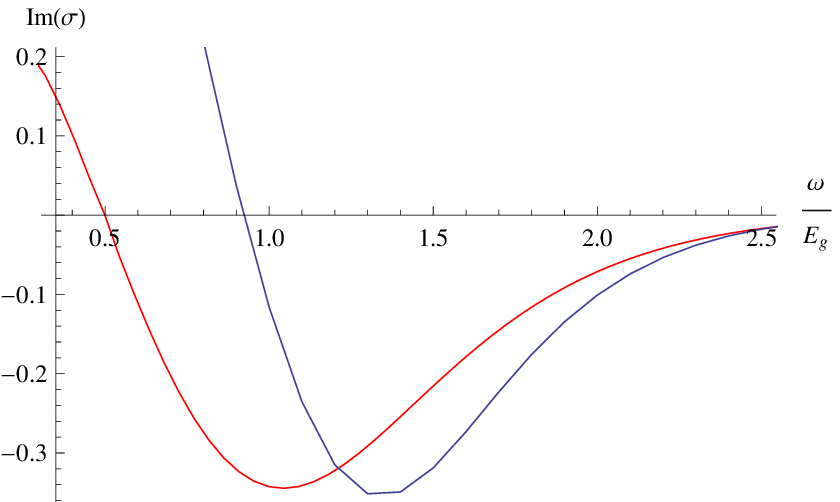}
\caption{Comparison of analytic and numerical results on the real and imaginary parts of the conductivity (left and right panel, respectively).} 
\label{figconda}
\end{figure}

We can look at the potential to determine the behavior of the conductivity near the critical line \eqref{eqcr}. We shall concentrate on the case $q^2=0$, in which the system simplifies considerably. Our discussion can be extended to other values of the coupling constant in a straightforward, albeit tedious, manner, but no new physical insight is gained.

\begin{figure}[ht!]
\includegraphics[width=.49\textwidth]{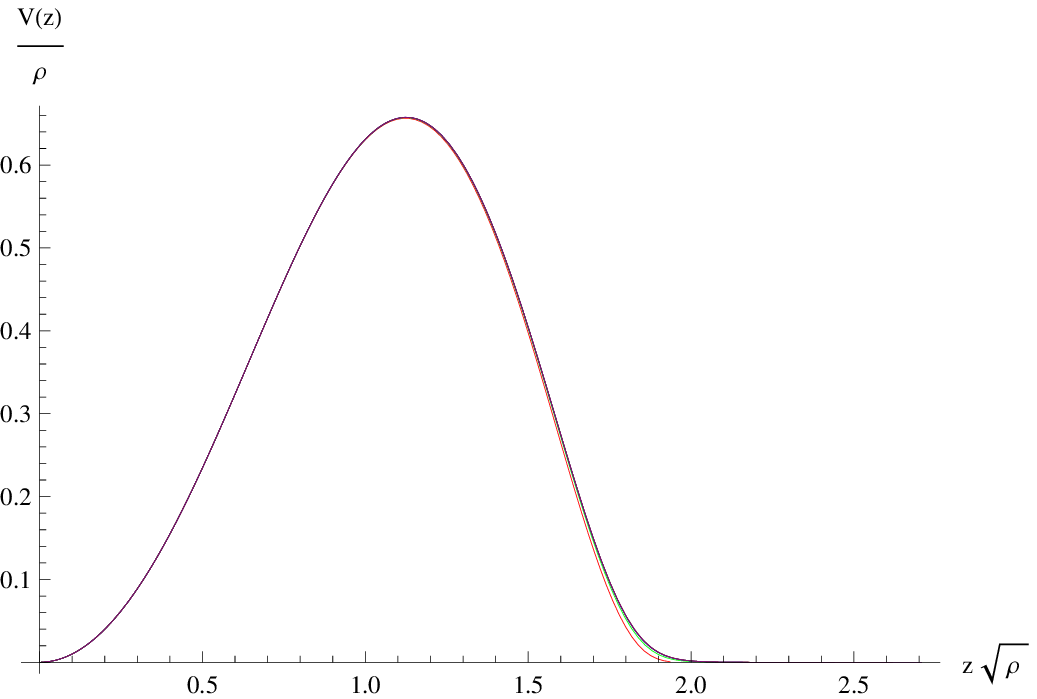}
\includegraphics[width=.49\textwidth]{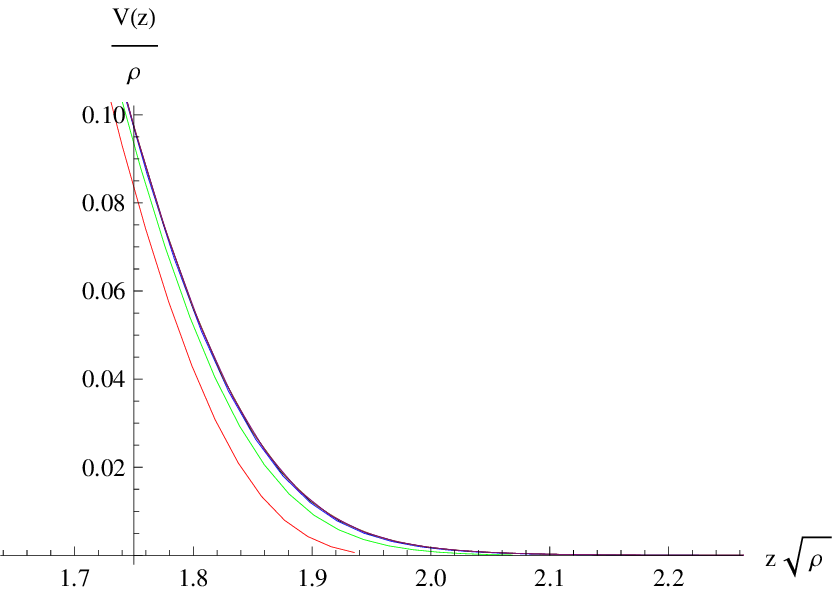}
\caption{Left panel: the potential determining the conductivity for $q=0$ at various temperatures below $T_c$. Right panel shows detail near the horizon.}
\label{fig5}
\end{figure}

Regardless of the value of $\Delta$, the potential reaches a universal shape and never becomes large enough for spikes to form at or near the BF bound, as seen previously in the probe limit \cite{Horowitz:2008bn,Horowitz:2009ij,Siopsis:2010pi}. The potential is plotted in figure \ref{fig5} below the critical temperature. After scaling $V$ by the charge density $\varrho$ and $z\to z\sqrt\varrho$, the curves at various temperatures $T\le T_c$ are almost indistinguishable.
Detail near the horizon (right panel of figure \ref{fig5}) shows the effect of lowering the temperature: the potential approaches zero faster at the horizon resulting in a longer tail.
Consequently, the conductivity also approaches a universal shape (figure \ref{figq0n}) below the critical temperature.

\section{Conclusion}
\label{sec6}

We discussed the properties of a hairy black hole near extremality corresponding (via a gauge/gravity duality) to a system of low ratio $T_c/\mu$ where $T_c$ is the critical temperature at which an instability arises and $\mu$ is the chemical potential. Using hypergeometric fucntions in the special case $m^2 =-2$, and trial functions in general, we calculated $T_c$ analytically for small charge of the scalar hair continuing analytically to negative values of $q^2$ as needed, down to $T_c=0$, which occurs at a critical value $q^2 = q_c^2$. Below $q_c^2$, there is no instability and no scalar hair develops. The critical point $q_c^2$ was expected to be given by \eqref{eqcr} as a function of the scaling dimension $\Delta$, which is of geometric origin, arising from the AdS$_2$ geometry of the horizon at extremality.
We showed that this is only true for $\Delta \ge \Delta_0$, where $\Delta_0 \approx 0.64$, whereas below this value, the true critical line is lower. As we approach the unitarity bound, the critical coupling remains negative, with $q_c^2 =0$ at $\Delta = 0.5$.
Among other things, this implies that for $q=0$, we have instability all the way down to the unitarity bound contrary to geometrical expectations.

Below the critical temperature, we calculated the condensate and the conductivity.
The energy gap in units of $T_c$ was found to diverge as we approached the critical line (i.e., $T_c\to 0$), and the conductivity as a function of frequency (appropriately normalized) was found to have a universal shape below the critical temperature. In particular, there is no strong dependence on the charge, which is different from the probe limit where there is a frequency gap that has an explicit $q$ dependence ($\omega_g \sim \langle q\mathcal{O}_\Delta \rangle^{1/\Delta}$).

Despite the mild dependence of physical quantities on the charge, it should be noted that the system of zero charge has an enhanced symmetry, since the phase of the scalar field decouples. It is therefore worth exploring its properties further, perhaps using a solution to the equations of motion with a non-trivial phase dependence, such as a vortex.



\acknowledgments

J.~A.\ acknowledges support from the Office of Research at the University of Michigan-Flint. J.~A.\ would also like to thank Colorado State University-Pueblo for their hospitality at the beginning of this work.
G.~S.~and J.~T. are supported in part by the Department of Energy under grant DE-FG05-91ER40627.

\end{document}